\documentclass{PoS}

\title{How to do a $\nu_{e} \rightarrow \nu_{\mu}$ measurement in a SK-like detector}

\ShortTitle{How to do a $\nu_{e} \rightarrow \nu_{\mu}$ measurement in a SK-like detector}

\author{\speaker{CHIZUE ISHIHARA}%
         \thanks{This work was supported in part by Global COE Program
(Global Center of Excellence for Physical Sciences Frontier), MEXT, Japan.}\\
        ICRR, University of Tokyo\\
        E-mail: \email{isihara@icrr.u-tokyo.ac.jp}}

\abstract{
In the future neutrino experiments,
a beta-beam, which can produce pure electron neutrino beam, is expected to 
achive precise measurement of the neutrino oscillation parameters.
In the $\nu_{\mu}$ appearance measurement of a beta-beam, 
a detector does not need to identify particle charge and thus,
a water Cherecov detector will be a candidate for the far detector.
In this paper, we study the expected signal detection efficiencies and background 
at the proposed beta beam facilities with the water Cherenkov detector.
In the estimation, we use the current simulation
and analysis tools developed for the Super-Kamiokande experiment.
Depending on the beta beam setups, the signal detection efficiencies
are found to vary from 36.4$\%$ to 75.3$\%$ in the standard $\nu_{\mu}$ search.
The major source of background was found to be neutral current pion production, 
and the fraction of the background increases with the mean energy of the neutrino beam.
}

\FullConference{10th International Workshop on Neutrino Factories, Super beams and Beta beams\\
                 June 30 - July 5 2008\\
                 Valencia, Spain}

\begin{document}

\section{Introduction}
Various neutrino experiments are in operation or in preparation to
measure neutrino oscillation parameters precisely.
These experiments are expected to provide the mass squared
differences or oscillation angles. However, the mass hierarchy of neutrinos or the CP phase
parameter of neutrino mixing still remains unkonwn.
In order to investigate these properties of neutrinos, it is essential
to have a much more intense and pure neutrino beam. Among the
proposed beam lines, the beta-beam has several unique features.
Since the beta-beam facility can produce pure electron neutrinos or anti-neutrinos 
from the decay of stored radio-active ion beams,
%it provides essentially background-free beam, i.e. enables us to measure 
%$\nu_{\mu}$ appearance in pure golden-channel ($\nu_{e} \rightarrow \nu_{\mu}$). 
it is not necessary to identify the charge
of leptons in the far detector, this will ease the requirements
of the far detector.
In this paper, we study the performance of the water Cherencov detector with a beta-beam
in light of experience gained from the Super-Kamiokande analysis.
\section{Neutrino Measurement in a Ring Imaging Water Cherenkov Detector}
\begin{figure}[h!]%%%%%%%%%%%%
\begin{center}
\includegraphics[height=9pc]{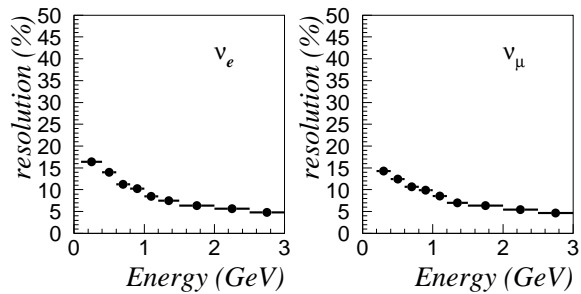}
%  \vspace{-1.5pc}
\caption{Resolution of neutrino energy as a function of the true neutrino energy for
1-ring, e-like, CCQE $\nu_{e}$ events(left) and 1-ring, $\mu$-like, CCQE $\nu_{\mu}$ events(right).}
\label{fig:enereso}
\end{center}
\end{figure}

%Super-Kamiokande (Super-K)~\cite{sk} is a 22.5 kt fiducial mass water Cherencov 
%detector located in the Kamioka mine, Gifu, Japan.
%The detector is optically separated into two concentric cylindrical regions.
%The detector is instrumented with 11,146 20-inch photomultiplier tubes(PMT)
%on over all inside of the tank for SK-I phase.
Super-Kamiokande(Super-K)~\cite{sk} is a 50 kt water Cherencov detector, which detects
the Cherencov ring image induced by charged particles and gamma rays.
Atmospheric neutrino oscillations have been established by the observation 
in Super-K~\cite{skatm}.  
The energy spectrum of neutrinos from the beta-beam facility
spreads from a few hundred MeV to a few GeV. Therefore,
it is possible to use the same method as used in the study of atmospheric
neutrinos. 
Since the beam direction is fixed, the incoming neutrino
energy is reconstructed as follows, by assuming 
charged current quasi-elastic(CCQE) interaction;
\begin{displaymath}
E_{\nu}^{rec} = 
\frac{m_{N} E_{\mu} - m^2_{\mu} / 2}{m_{N} - E_{\mu} + P_{\mu}\cos\theta_{\mu}} ,
%\Label{eq:rec}
\end{displaymath}
where $m_{N}$ is the nucleon mass, $E_{\mu}$ is the muon energy, 
$m_{\mu}$ is the muon mass and $P_{\mu}$ is the muon momentum. 
The actual steps to search for CCQE events is
as follows~\cite{skatm}: (1) search for an event with a single Cherenkov ring of
a lepton produced by neutrino charged current interaction,
(2) classify the ring into two categories, e-like and $\mu$-like, 
using the photon distribution of the ring pattern, and
(3) reconstruct the momentum and direction of the lepton using
the observed ring image.
%\begin{enumerate}
%\item search for an event with single Cherenkov ring of
%      a lepton produced by the neutrino charged current interactions,
%\item classify the ring into two categories, e-like and $\mu$-like, 
%      using the photon distribution of the ring pattern, and
%\item reconstruct the momentum and direction of the lepton using
%      the observed ring image.
%\end{enumerate}
%
The resolution of neutrino energy for CCQE events is shown in Fig.~\ref{fig:enereso}.
%%%%%%%%%%%%%%%%%
\section{Analysis}
The aim of this study is to estimate the detection efficiencies and the background
in the search for $\nu_{\mu}$ appearance with a beta-beam and water Cherencov detector.
As the tools to study these items, we use available simulation
and analysis programs from the Super-Kamiokande experiment because their
performances have been verified in the past atmospheric
neutrino and accelerator neutrino studies.
\subsection{Neutrinos from the beta-beam facility}
The properties of the neutrino beam from a beta-beam facility is determined by 
the type of ion and its relativistic $\gamma$. 
It is possible to estimate the energy spectrum of the neutrino beam very precisely 
because the kinematics of beta decay is very well understood.
The mean energy of the neutrino beam needs to be adjusted to maximize
the sensitivities for neutrino observation. Usually, the peak energy of the neutrino beam
is adjusted to the oscillation maximum, which is determined by the baseline
together with the energy of the neutrinos.
In this study, two baseline distances (L) are selected as described in~\cite{bb}.
One of them is 130km corresponding to the distance from CERN to Frejus, and
the other is 700km corresponding to the distance from CERN to Gran-Sasso.
%A low energy beam, peaked at 0.4GeV, matches the distance 
%from CERN to Frejus, L = 130km.
%In this option, the appropriate $\gamma$ (= 100) could be achived with the present CERN-SPS
%by using He(Ne) ions for antineutrinos(neutrinos) production.
%A high energy beam, peaked at 1-1.5GeV, could reach the second oscillation 
%maximum at L $\sim$ 700km, matching the distance, for example, 
%between CERN and Gran-Sasso. Refurbished SPS could
%be achived $\gamma = 350$ for both He and Ne.
%It is proposed that the beam peaked at 0.4GeV (LE beam), matching first oscillation 
%maximum and the beam peaked at 1-1.5GeV (HE beam), matching second oscillation maximum. 
For the shorter baseline case, the oscillation maximum is around $0.4GeV$, and
for the other case, it is around $1.5GeV$. 
So here the neutrino beam of peak
energy of $\sim0.4GeV$ is referred to as the 'LE beam' configuration, and
peak energy of $\sim1.5GeV$ as the 'HE beam' configuration.
Also both $\nu_e$ and $\bar{\nu_e}$ beams are necessary to study
CP violation and the mass hierarchy.
These four sets of ion and $\gamma$ combinations have been identified as
the candidate configurations, as described in~\cite{bb}.
%These beams are available for both $\nu_{e}$ and $\bar{\nu_{e}}$.
%So we assume 4 beta-beam setups, LE $\nu_{e}$, LE $\bar{\nu_{e}}$, HE $\nu_{e}$ and
%HE $\bar{\nu_{e}}$, showing left panel of fig.68 in ~\cite{bb2}.
%
\subsection{Event selection criteria and signal efficiencies}
\begin{figure}[h!]%%%%%%%%%%%%
\begin{center}
\includegraphics[height=10pc]{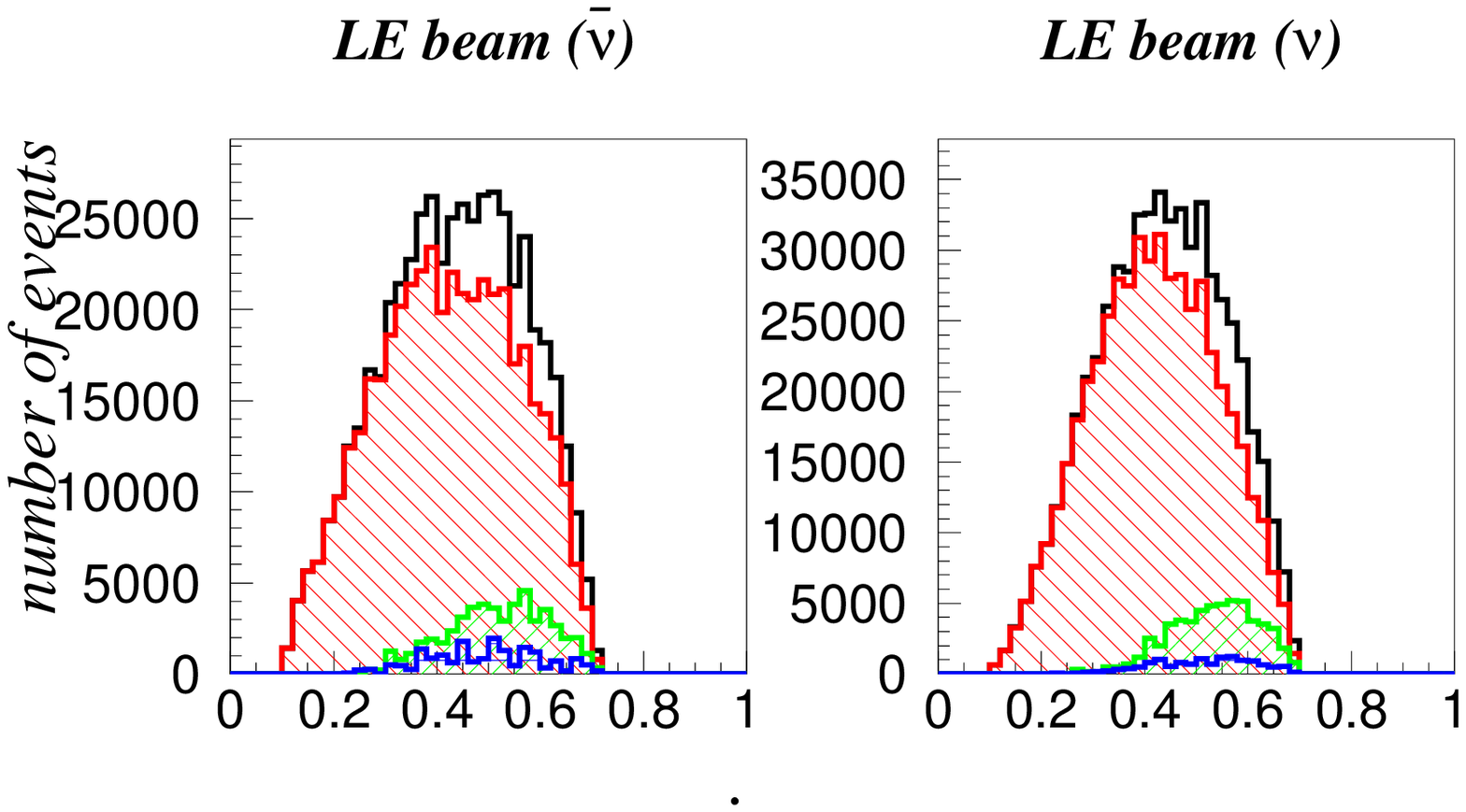}
\includegraphics[height=10pc]{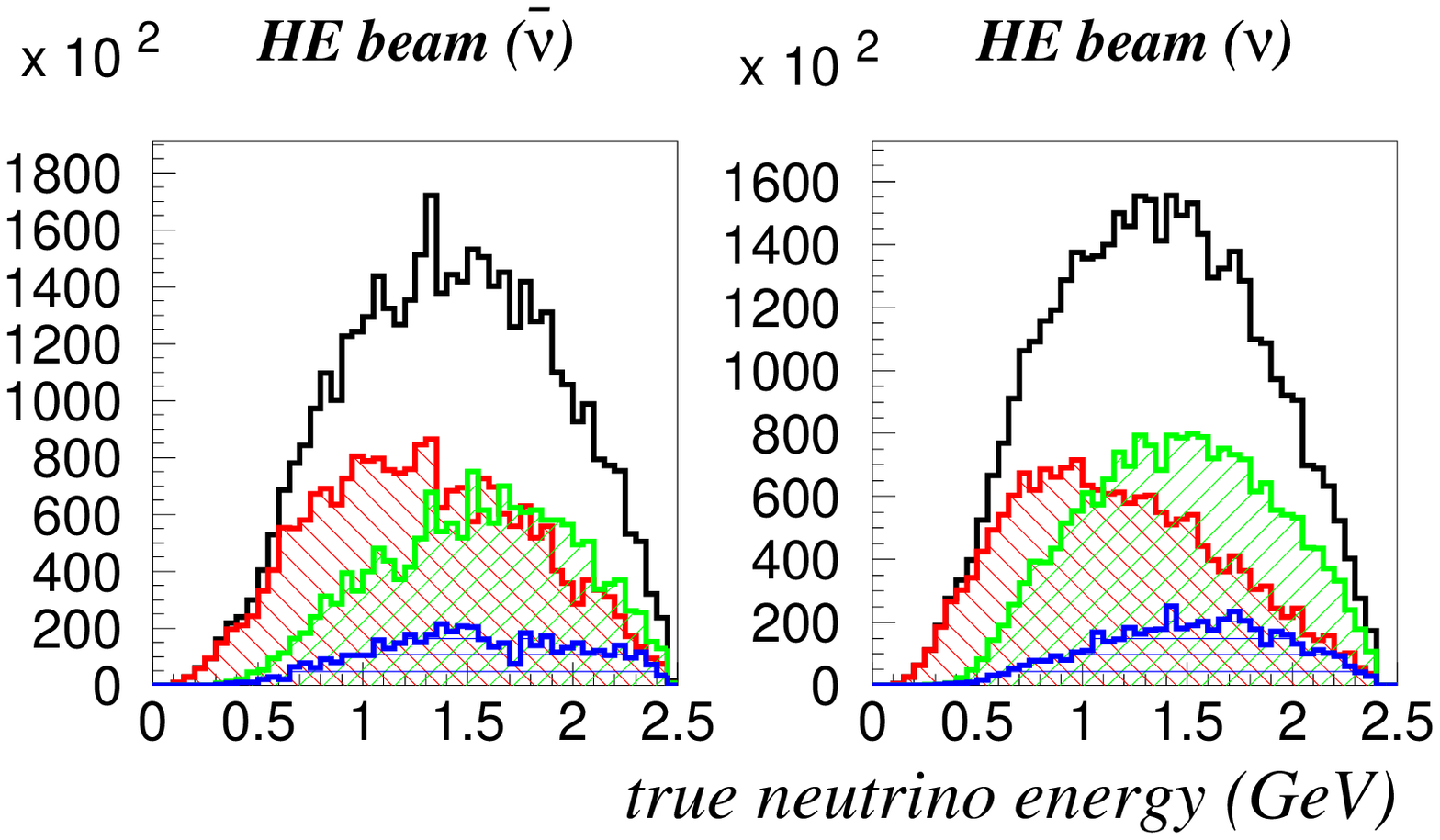}
\includegraphics[height=3pc]{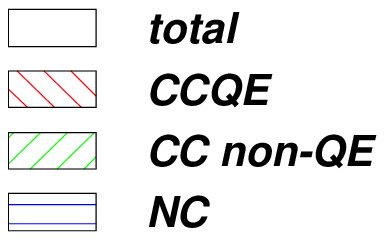}
%  \vspace{-1.5pc}
\caption{The rate of neutrino interactions in water Chrencov detector for each beam setup.}
\label{fig:intrate}
\end{center}
\end{figure}

At first, we apply standard event selection to eliminate the
cosmic ray muons and the very low energy events. The selection
criteria are that there is no activity in the outer detector(FC event),
the reconstructed vertex is in the fiducial volume(FV), and
the electron equivalent energy(evis) is larger than 30MeV.
The neutrino energy spectra for each beam configuration after
this selection are shown in Fig.~\ref{fig:intrate}. It should be noted that
events other than CCQE interactions
will be dominant in the HE beam configurations.

In the following analysis, the CCQE events need to be selected as discussed before.
Because the Cherenkov threshold of the proton in a water cherenkov
detector is 1.1 GeV, only the lepton is identified as a clear ring
in this energy range. Therefore, an event with a single ring is selected.
Ring candidates are searched for based on the Hough transformation method~\cite{ring}
and the number of rings are determined by the likelihood method.
%Let's discuss the selection efficiency of CC events by Super-K event selection algorithm.
%1-ring event is selected to use CCQE event which momentum can be reconstructed correctly.
%Ring candidate serching is base on Hough transformation method and test the ring
%by the likelihood method. The efficiency for idetifying CCQE events as 1-ring 
%is 94.2$\%$ for atmospheric neutrinos. 
%After 1-ring selection, the event fraction from beta-beam neutrino is Fig..~\ref{fig:ring} 
%The fraction goes down with increasing neutrino energy, beacuse 
%CC non-QE interactions is getting larger over 1GeV.
The selected ring is classified into two types~\cite{skatm},
e-like and $\mu$-like, by using the difference of the shape of the ring.
The misidentification probability is about 1$\%$
for both atmospheric $\nu_{e}$ and $\nu_{\mu}$ of CCQE events.
The $\mu$-like selection probability in the beta-beam neutrino sample after 1-ring selection 
is about 90$\%$ for <400MeV, 95$\%$ for >400MeV energy region.
However, the events selected as a $\mu$-like event sample
contain the charged pions
because the ring shape of the charged pions are similar
to the muon rings, rather than the electron rings. 
So the events except CCQE interaction are contaminated. 
In order to eliminate those charged pion events,
events with 1 decay electron are selected. Because the interaction
probability of charged pions in water is quite large,
a large fraction of the charged pions interact before decaying.
%that is, reduce the events other than CCQE interaction. 
The efficiency of detecting decay-electrons in Super-K 
is 80$\%$ (63$\%$) for $\mu^{+}$ ($\mu^{-}$).
The summary table of event selection efficiencies is shown in Tab.~\ref{tab:eff}.
%%%%%%%%%%%%%%%%%
\begin{table}[h!]%%%%%%%%%%%%
\begin{center}
\includegraphics[height=10pc]{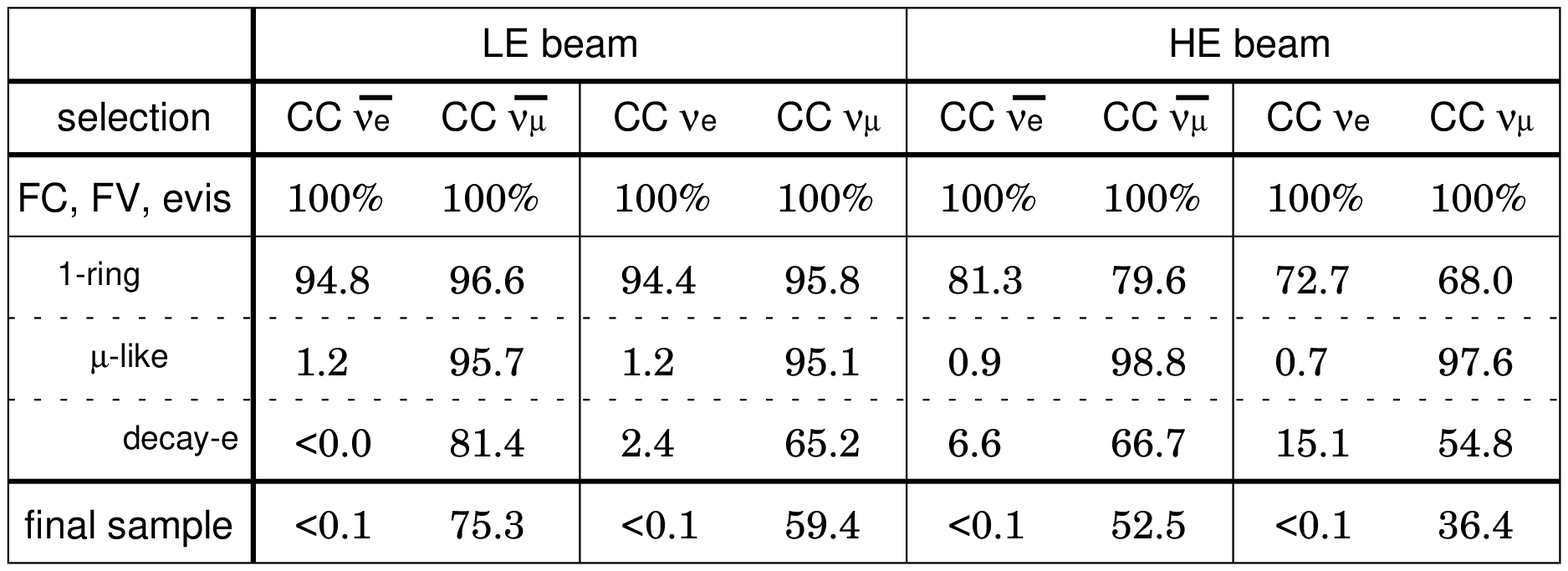}
%  \vspace{-1.5pc}
\caption{Summary of selection efficiencies of CC events for each beam set. 
After standard selection(FC, FV and evis>30MeV), the numbers in each selection step show 
the probabilities after the previous step.}
\label{tab:eff}
\end{center}
\end{table}%%%%%%%%%%%%%%%%%
\subsection{Background events}
\begin{figure}[h!]%%%%%%%%%%%%
%\begin{center}
\begin{minipage}{0.38\hsize}
 \includegraphics[width=1.\textwidth]{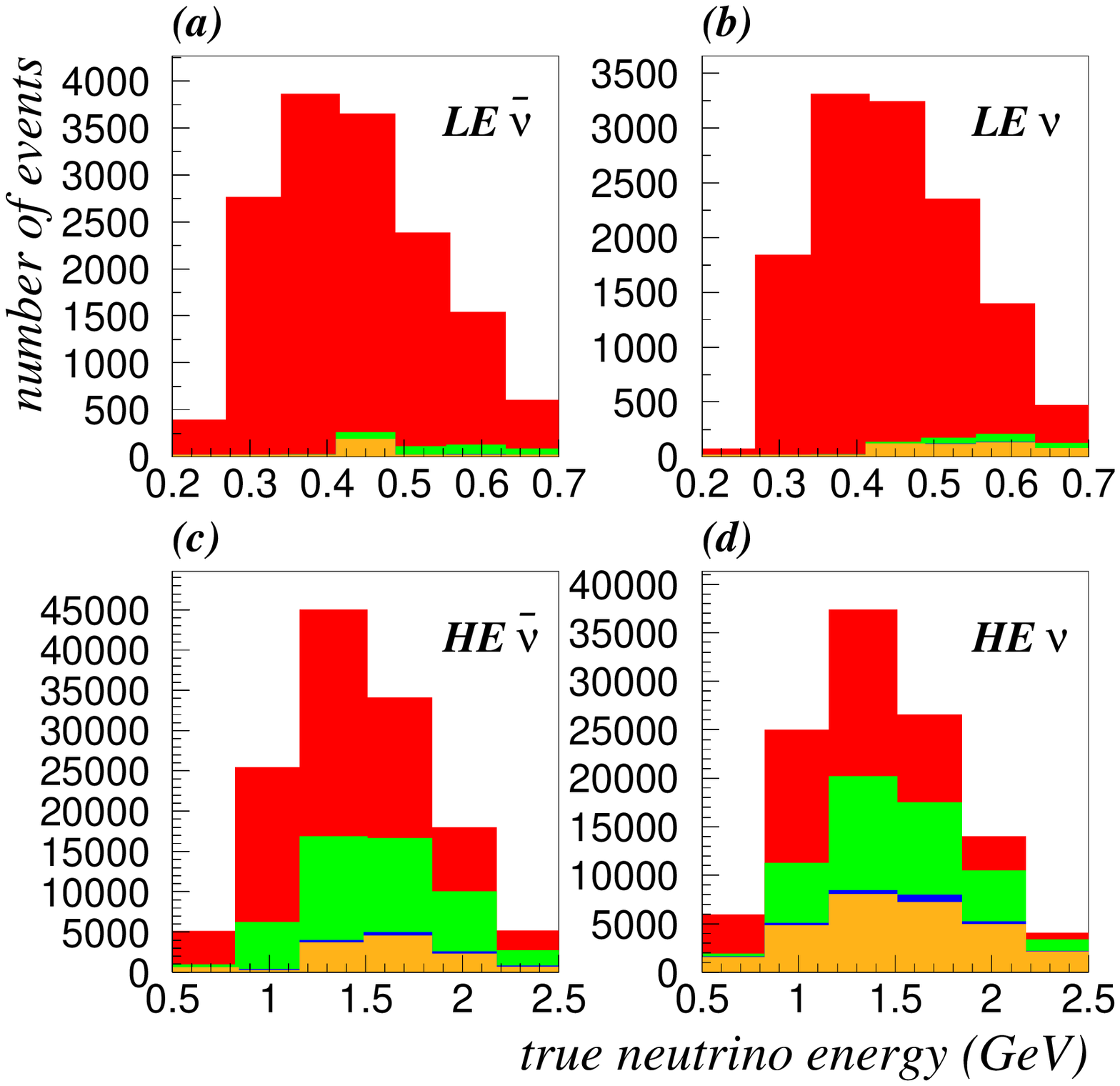}
 \caption{The final sample neutrino true energy distributions for each beam type,
 in case of $\sin^{2}2\theta_{13} = 0.15$. The different event types are shown in 
 different colors as shown right side.}
 \label{fig:enedis1}
\end{minipage}
\hspace{2mm}
\begin{minipage}{0.42\hsize}
 \includegraphics[width=.48\textwidth]{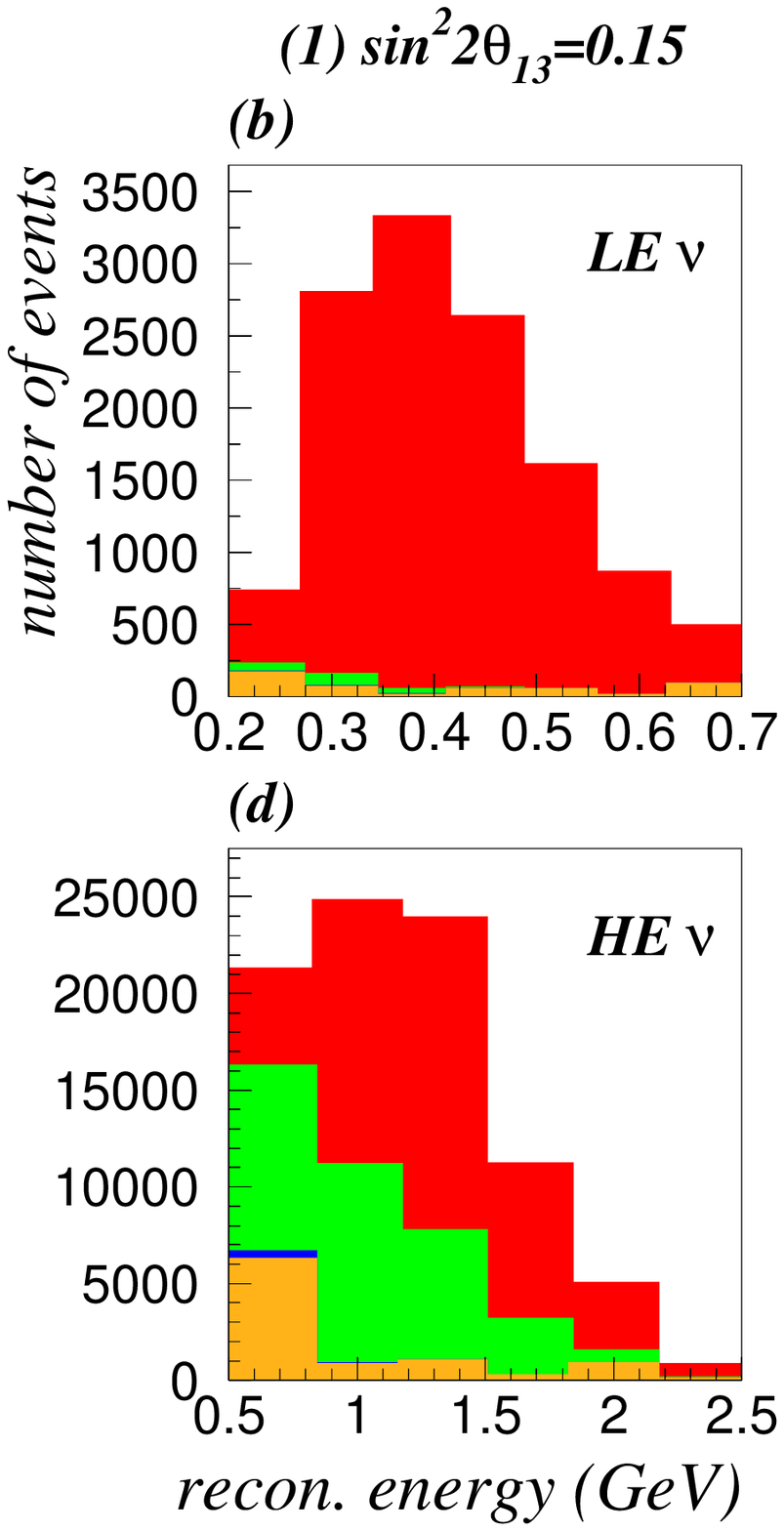}
 \includegraphics[width=.48\textwidth]{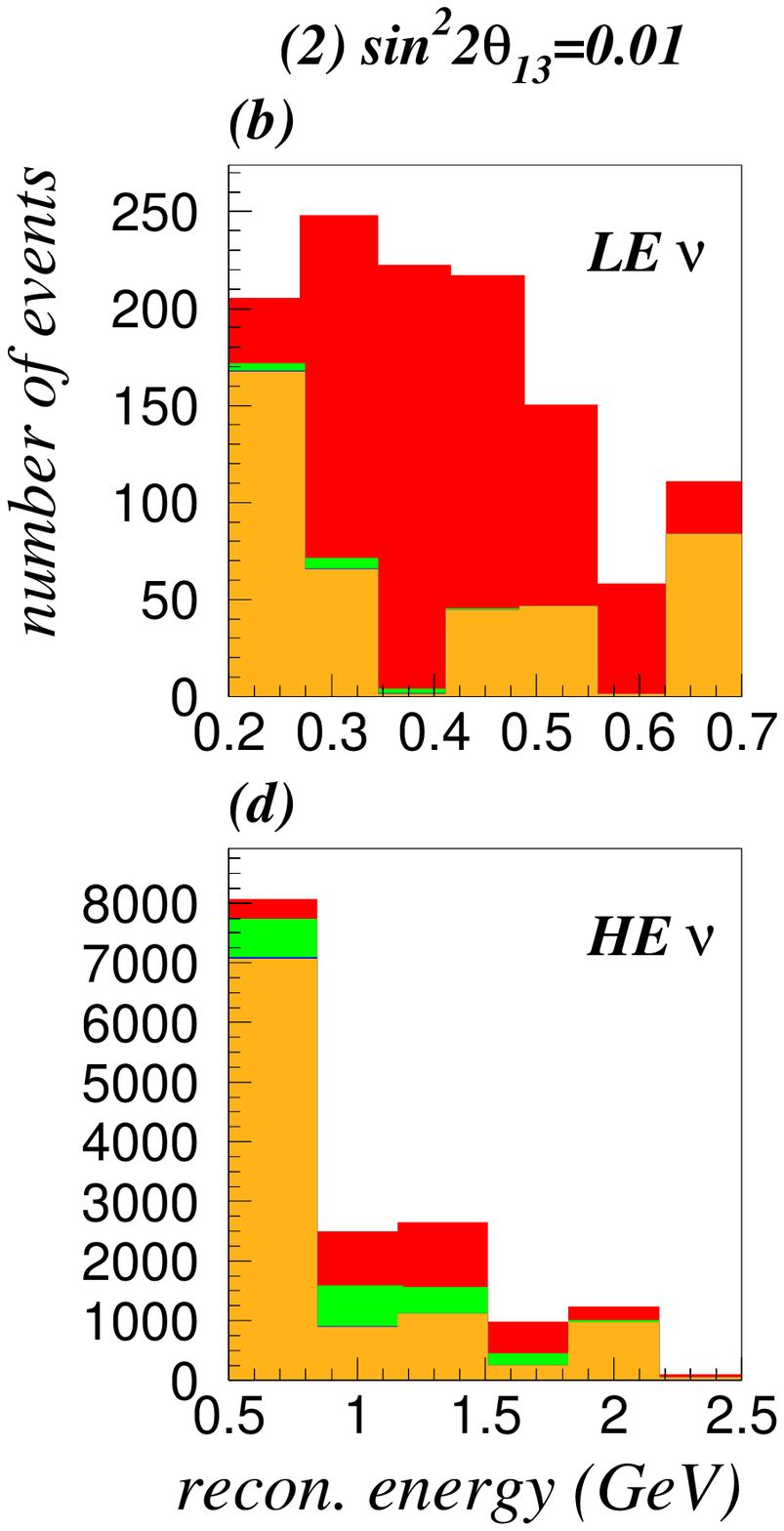}
 \caption{(1) are the reconstructed energy distributions in case of $\sin^{2}2\theta_{13} = 0.15$.
 (2) are in case of $\sin^{2}2\theta_{13} = 0.01$.}
 \label{fig:enedis2}
%\end{center}
\end{minipage}
\hspace{1mm}
\begin{minipage}{0.15\hsize}
  \includegraphics[width=1.\textwidth]{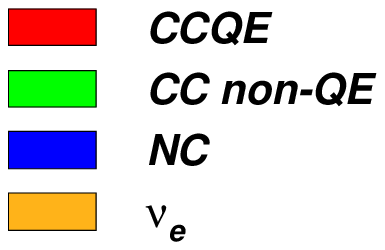}
\end{minipage}
\end{figure}

Major background events in the search for $\nu_\mu$ appearance
signal are produced by NC interactions, beacause pions are selected as 
$\mu$-like events.
The cross-section of the NC pion production is larger of
higher neutrino energies and thus,
the fraction of the backgournd
events get higher in the HE beam as shown in Fig.~\ref{fig:enedis1}-(c),(d),
in which true paramteres are assumed: $\sin^22\theta_{13}=0.15$, $\sin^2\theta_{23}=0.5$, 
$\Delta m^2 = 2.5 \times 10^{-3} eV^{2}$ and $\theta_{12}=0$ (no matter effect).
Moreover, the momentum of the pions are rather low because most of
the pions are produced via decay of resonances. 
Therefore,
the reconstructed energy using the misidentified pions 
is rather low and not sensitive to the energy of the parent neutrino.
%contaminate in the oscillation signal.
%BG in $\nu_{\mu}$ appearance signal are NC and $\nu_{e}$ events.
%The left figures at Fig.~\ref{fig:enedis} show the distributions of neutrino true energy,
%in case $\sin^22\theta_{13}=0.15$, $\sin^2\theta_{23}=0.5$, 
%$\Delta m^2 = 2.5 \times 10^{-3} eV^{2}$ and $\theta_{12}=0$ (no matter effect).
%Especially, much $\nu_{e}$ BG are contaminated in high energy beam. 
%$\nu_{e}$ NC events are significant BG because pions from NC single-$\pi$ production
%make $\mu$-like. 

But, in case of the reconstructed energy distributions (Fig.~\ref{fig:enedis2}-(1)), 
the NC events are peaked at lower energy because of the different kinamatics 
between QE and NC events.
It is therefore easy to distinguish these background events from the signal.
For the $\sin^{2}2\theta_{13} = 0.01$ case (Fig.~\ref{fig:enedis2}-(2)), 
it is difficult to see $\nu_{\mu}$ signal, especially in the HE beam(Fig.~\ref{fig:enedis2}-(2-d)). 
%%%%%%%%%%%%%%%%%
\section{Conclusion}
$\nu_{\mu}$ appearance measurement by using beta-beams and 
water Cherencov detectors is very promising as a future neutrino 
oscillation experiment.
In these experimental configurations, the signal detection efficiencies of 
a water Cherencov detector are expected to be
75.3$\%$(59.4$\%$) for $\bar{\nu_{\mu}}$($\nu_{\mu}$) from the LE beam 
and 52.5$\%$(36.4$\%$) for $\bar{\nu_{\mu}}$($\nu_{\mu}$) from the HE beam.
%The small fraction of background in the LE beam is expected, and For HE beams, CC non-QE, NC
%and $\nu_{e}$ events are contaminated. 
Meanwhile, becasue the interactions other than CCQE interaction become dominant at high energy,
the rate of background increases in the HE beam configurations.
Mainly, misidentification of the pion's rings from NC $\nu_e$ pion production 
is the cause of the background.
%If improvements are made to the PID selection,
%$\nu_e$ pion production can be reduced.
% and it could be helpful
%for pion's rings of NC$\nu_e$ interaction to separate from muon's.
%It is possible to distinguish NC events by using the reconstructed neutrino energy.
%
%
%\section*{Acknowlegements}
%I would like to thank Pilar Hernandez
%for this opportunity.
%I would like also to thank Prof. T. Kajita, Y. Hayato 
%and all Super-Kamiokande collaborators.
%
%
%%%%%%%%%%%%%%%%%%%%%%%%%%%%%%%%%%
%% thebibliography environment %%
%%%%%%%%%%%%%%%%%%%%%%%%%%%%%%%%%

%%%%%%%%%%

\end{document}